\documentstyle[12pt,psfig]{article}
\newcommand{\mathd}[1]
{ \begin{equation} #1 \end{equation}}

\newcommand{\mathld}[1]
{ \begin{eqnarray} #1 \end{eqnarray}}

\newcommand{\tr}[0] { {\rm Tr}}

\begin{document}

\hfill PUPT-1551

\hfill {\tt hep-th/9507062}

\hfill July 1995

\begin{center}

{\bf Matrix Model Approach to $d>2$ Non-critical Superstrings}

\vspace{1ex}

Akikazu Hashimoto and Igor R.~Klebanov\\

\vspace{1ex}

{\it Department of Physics\\
          Princeton University\\
          Princeton, NJ 08544}\\

\end{center}

\begin{abstract}

We apply light-cone quantization to a $1+1$ dimensional supersymmetric
field theory of large N matrices.  We provide some preliminary
numerical evidence that when the coupling constant is tuned to a
critical value, this model describes a 2+1 dimensional non-critical
superstring.

\end{abstract}

Matrix models have been remarkably successful \cite{MatrixReview} in
providing us with an understanding of non-critical bosonic string
theories in dimensions less than or equal to two.  Their extension to
dimensions greater than two, however, runs into a number of
difficulties, some of them related to the appearance of tachyons in
the bosonic string spectrum. Quantization in the infinite momentum
frame is a promising approach to these models, where strings emerge as
bound states of more elementary ``string bits''
\cite{OldThorn,KlebanovSusskind88}.  Studies of the $c=2$ matrix model
using this technique have indeed revealed the expected tachyon problem
\cite{LightconeC=2,MoreC=2}.

While the bosonic strings are unstable above two dimensions, one
expects this instability to be cured by target space
supersymmetry. Full space-time supersymmetry requires that the string
theory be ten-dimensional (with a possibility of subsequent
compactification).\footnote{A successful construction of critical
superstrings from{} string bit models has been found recently
\cite{BergmanThorn95}.}  It is well known, however, that in $c+1$
dimensional non-critical string theories the Liouville coordinate
enters on a different footing from{} other dimensions, so that there is
no full Poincare invariance. Therefore, one cannot demand a full $c+1$
dimensional supersymmetry. The most one could require is the $c$
dimensional ``space supersymmetry'' which was considered in ref.
\cite{SK90}. While this work utilized the NSR approach to the
super-Liouville theory, it is interesting to ask whether an equivalent
Green-Schwartz formulation exists \cite{SiegelGS,SiegelLiouville}. An
even more ambitious goal is to construct matrix models which describe
non-critical superstring theories. In this paper we make a small step
in this direction by formulating and numerically studying a
two-dimensional supersymmetric matrix field theory which is expected
to describe $2+1$ dimensional non-critical superstrings.

Our model is a higher-dimensional generalization of the
Marinari-Parisi model (the supersymmetric quantum mechanics of a
large-$N$ hermitian matrix) \cite{MP90}.  It is not entirely clear
whether the continuum limit of the Marinari-Parisi model describes a
theory with target space supersymmetry; in fact, the conventional
understanding is that this model provides an alternative description
of the bosonic $c=0$ string \cite{MP90,CP94}.  There are some signs,
however, of remaining target space supersymmetry
\cite{MPFollowup}. Since one-dimensional supersymmetry is quite
trivial, we hope that going one dimension higher will reveal a more
interesting structure.  The price we have to pay is that the model is
no longer exactly solvable. Luckily, using the techniques of
light-cone quantization, we can still set up a scheme for calculating
the spectrum of the matrix model in the large-$N$ limit. Our approach
is largely an extension of the program of \cite{LightconeC=2,MoreC=2},
with the advantage that the supersymmetry cures the sickness of the
model.

The $c=2$ matrix model studied in   \cite{LightconeC=2}
is defined as a two-dimensional field theory
with the euclidean action
$$ S=\int d^2 x \tr \left ({1\over 2} (\partial_\mu \phi)^2+
{1\over 2} m^2 \phi^2-
{1\over 3\sqrt{N}}\lambda \phi^3 \right )\ ,
$$
where $\phi(x^0, x^1)$ is an $N\times N$ hermitian matrix field.  The
connection of this model with triangulated random surfaces follows, as
usual, after identifying the Feynman graphs with the graphs dual to
triangulations.  At the leading order in $N$, we obtain a sum over the
planar triangulated random surfaces embedded in two dimensions.  If
the tadpole graphs are discarded, as they should be because they do
not correspond to good triangulations, then the entire perturbative
expansion is finite.  This is similar to what we find in the matrix
models for $c\leq 1$.  Therefore, it is sensible to look for a
singularity in the sum over the planar graphs as $\lambda$ approaches
some critical value $\lambda_c$. For the $c=1$ model the spectrum of
energies becomes continuous at $\lambda=\lambda_c$. This is because we
are really dealing with a $1+1$ dimensional string theory whose
center-of-mass mode satisfies \cite{LiouvilleDispers} $$ E^2 =
p_\phi^2 $$ where $ p_\phi$ is the Liouville momentum which possesses
continuous spectrum. By analogy one might expect that in the $c=2$
model the spectrum of $M^2= 2P_+ P_-$ becomes continuous at
$\lambda=\lambda_c$.  Light-cone quantization of the model revealed,
however, that the mass-squared of the lowest state becomes negative
for sufficiently large $\lambda$ and, in fact, tends to $-\infty$ as
$\lambda\rightarrow\lambda_c$ \cite{MoreC=2}.  Although it is still
possible that a continuum of states is forming around $M^2=-\infty$,
it was difficult to draw definite conclusions from{} numerical studies
of the spectrum. Further evidence for a phase transition at some
critical value of the coupling was given \cite{Dalley94} based on an
approximate identification of the $c=2$ model with a one-dimensional
spin chain. However, the fact remains that the corresponding continuum
theory, even if it exists in some sense, is a very sick theory
containing a tachyon.

In this article, we attempt to cure this sickness by considering a
$1+1$ dimensional large-$N$ hermitian matrix field theory with $(1,1)$
supersymmetry
$$  [P_+,P_-] =    [P_\pm,Q_\pm] =   \{ Q_+,Q_- \}  =  0\ ,$$
$$  Q_+^2  = P_+,\ \ \  Q_-^2  =  P_-\ .$$
{}From{} the above algebra it follows that the operators $P_+$ and $P_-$
are positive definite, since they are squares of hermitian
supercharges $Q_+$ and $Q_-$. This implies that $M^2 = 2P_+P_-$ is
also positive definite. As we tune the parameters, we expect the
theory to become effectively $2+1$ dimensional due to the appearance
of the Liouville mode which makes the spectrum of $M^2$
continuous. Furthermore, we expect the mass-squared of the lightest
state to vanish at this critical point, since the appearance of new
massless states is typically a signature of the phase transition.
Indeed, the continuous spectrum with a vanishing mass gap was observed
in the $c=1$ \cite{LiouvilleDispers} and the Marinari-Parisi models
\cite{MP90}.  In this article, we present some evidence that such
massless continuum of states does appear in the spectrum of $1+1$
dimensional supersymmetric matrix field theory.

The simplest matrix field theory with the above supersymmetry is given
by the action
$$ S = \frac{1}{4} \int d^2x d^2 \theta \tr \left[\bar{D} \Phi D \Phi +
W(\Phi)\right]$$
where $\Phi$ is a (1,1) matrix superfield
$$\Phi_{ij} = \phi_{ij} + \bar{\theta}\Psi_{ij} + \bar{\theta}\theta F_{ij},$$
and $\Psi_{ij}$ is a matrix whose elements are two-component spinors
$$\Psi_{ij} = \left[
\begin{array}{c}
\Psi_{-} \\
\Psi_{+}
\end{array}
\right]_{ij}.
$$
In light-cone quantization only $\Psi_-$ turns out to be
dynamical. $D$ is the covariant derivative
$$D = \frac{\partial}{\partial \bar{\theta}} - i \Gamma^\mu \theta
\partial_\mu.$$
where $\Gamma^\mu$ are the two dimensional Dirac matrices in the
Majorana representation
$$\Gamma^0 = \left[
\begin{array}{cc}
   & -i \\
 i &    \\
\end{array}
\right],\ \ \
\Gamma^1 =
\left[ \begin{array}{cc}
   & i \\
 i &    \\
 \end{array}
\right]. $$
We consider the simplest superpotential $W(\Phi)$  given by
$$W(\Phi) = \frac{1}{2} \mu \Phi^2 - {\lambda\over
3\sqrt N} \Phi^3.$$
The Feynman graphs which arise from{} the perturbative expansion of the
path integral generate triangulated random surfaces embedded in the
superspace with coordinates $(x^0, x^1, \theta^1,
\theta^2)$. We expect that, as $\lambda$ is tuned to some critical value
$\lambda_c$, the size of a typical graph diverges so that the world
sheet continuum limit may be defined.  The dimension two terms in the
world sheet action are fixed by the target-space supersymmetry (they
may also be inferred from{} the superspace propagator),
\mathd{
S_{\rm world\ sheet} = \int d^2\sigma \sqrt{h} h^{\alpha \beta}
\left[z_1
\left( \partial_\alpha x^\mu - i\bar{\theta} \Gamma^\mu \partial_\alpha \theta
\right)
\left( \partial_\beta x_\mu - i\bar{\theta} \Gamma_{\mu} \partial_\beta \theta
\right)
+i z_2 \partial_\alpha \bar{\theta} \partial_\beta \theta \right],
\label{eq.worldsheet}}
where $z_1$ and $z_2$ are normalization constants, and the metric
$h_{\alpha \beta}$ is regarded as a dynamical variable.  This action
resembles the Green-Schwartz action.  However, it has no Wess-Zumino
term or manifest $\kappa$-symmetry.  Nevertheless, if we find anything
non-trivial from{} our model, it is a kind of non-critical
Green-Schwartz superstring.

In terms of component fields, the matrix model action is written
\mathd{
S = \int d^2x \tr \left[ \frac{1}{2} (\partial_\mu \phi)^2 +
\frac{1}{2} \bar{\Psi} i\partial\!\!\!/ \Psi - \frac{1}{2} V^2(\phi) -
\frac{1}{2} V'(\phi)\bar{\Psi}\Psi\right] \label{eq.component.action}
}
where $V(\phi) = W'(\phi) = \mu \phi - {\lambda\over\sqrt N} \phi^2$.
As such, it could also be thought of as a matrix analogue of the model
considered in \cite{WittenOlive} with zero central charge.  The above
Lagrangian contains cubic and quartic interaction terms which depend
linearly and quadratically on the coupling constant $\lambda$.  One
could now proceed to enumerate the states and determine the matrix
elements of $P_-$ as was done in theories without supersymmetry
\cite{LightconeC=2,MoreC=2}.  Instead, we prefer to determine the
matrix elements of supercharges for the enumerated states, as was
recently proposed in \cite{MSS95}. This has a number of advantages.
Firstly, the supercurrent contains only terms up to cubic order in
fields, and depends only linearly on the coupling constant $\lambda$.
Secondly, the supercharge is a more fundamental dynamical object,
being the square root of the Hamiltonian.  We find it particularly
convenient to consider the combination $I = \sqrt{2} i Q_+ Q_-$.  This
operator is bosonic, commutes simultaneously with $P_+$ and $P_-$, and
its square is equal to $M^2$.  $I$ contains more information than
$M^2$ because it fixes the ambiguity of sign when one takes the square
root.  When we search for evidence for the continuous spectrum, this
fact will be rather useful.

The action is invariant under the standard supersymmetry transformation
$$ \delta \phi_{ij} = \bar{\epsilon}\Psi_{ij}$$
$$ \delta \Psi_{ij} = -i \Gamma^\mu \epsilon \partial_\mu \phi_{ij}. $$
The supercurrent can easily be shown to be
$$J^\mu = \frac{1}{\sqrt{2}}\tr \left[ (\partial\!\!\!/\phi )\Gamma^\mu\Psi + i
V(\phi)\Gamma^\mu \Psi \right]. $$
The light-cone components of the supercharges are
$$Q_- = \int dx^- : \tr \left[ \sqrt{2} \left( \partial_- \phi \right)
\Psi_-\right] :$$
$$Q_+ = \int dx^- :\tr \left[  V(\phi) \Psi_- \right ] :.$$
Expanding in normal modes\footnote{The factor of $1/\sqrt{2}$ appears
due to the unusual normalization of the kinetic term for $\Psi$ in
(\ref{eq.component.action}).}
$$\phi_{ij}(x^-) =
\frac{1}{\sqrt{2\pi}} \int_0^\infty \frac{dk_-}{\sqrt{2k_-}}\left(
a_{ij}(k_-) e^{-ik_-x^-} + a_{ji}^{\dagger}(k_-)e^{ik_-x^-}\right)$$
$$\frac{1}{\sqrt{2}}\Psi_{ij}(x^-) = \frac{1}{\sqrt{2\pi}}
\int_0^\infty \frac{dk_-}{\sqrt{2}} \left(
 b_{ij}(k_-) e^{-ik_-x^-} + b_{ji}^{\dagger}(k_-)e^{ik_-x^-} \right),$$
the supercharges become
\mathld{
iQ_- & = & \int_0^\infty dk \sqrt{k} \left(
b_{ij}^{\dagger}(k) a_{ij}(k)
 - a_{ij}^{\dagger}(k)
b_{ij}(k)\right)\nonumber \\
\sqrt{2} Q_+ & = & \int_0^\infty dk \frac{1}{\sqrt{k}}\left(
b_{ij}^{\dagger}(k) a_{ij}(k)
+ a_{ij}^{\dagger}(k) b_{ij}(k)\right) \nonumber \\
& & - {y\over\sqrt N} \int dk_1 dk_2 \left\{\rule{0in}{3ex} \right.\nonumber \\
& &\ \ \ \ \ \ \ \frac{1}{\sqrt{k_1 k_2}}
\left[
a_{ij}^{\dagger}(k_1) a_{jk}^{\dagger}(k_2) b_{ik}(k_1+k_2) +
b_{ik}^{\dagger}(k_1+k_2)a_{ij}(k_1)a_{jk}(k_2)
\right]
\nonumber \\
& &\ \ \ \ \ \ \ \frac{1}{\sqrt{k_1(k_1+k_2)}}
\left[a_{ij}^{\dagger}(k_1)b_{jk}^{\dagger}(k_2)a_{ik}(k_1+k_2)
+ a_{ik}^{\dagger}(k_1+k_2) a_{ij}(k_1) b_{jk}(k_2)\right] \nonumber \\
& &\ \ \ \ \ \ \ \frac{1}{\sqrt{k_2(k_1+k_2)}}
\left[b_{ij}^{\dagger}(k_1)a_{jk}^{\dagger}(k_2)a_{ik}(k_1+k_2)
+ a_{ik}^{\dagger}(k_1+k_2) b_{ij}(k_1) a_{jk}(k_2)\right] \left.
\rule{0in}{3ex} \right\}.  \nonumber
}
We have dropped the subscript ``minus'' on $k$ for brevity, set
$\mu=1$, and introduced the dimensionless coupling parameter $y =
\lambda / (2\sqrt{\pi} \mu)$.  The theory may be regulated in the infrared
by compactifying the $x^-$ direction and imposing periodic boundary
conditions on the fields.  The momentum $k$ then takes on discrete
values $n/L$, and the integrals above are replaced by sums.  For
massive particles we may discard the $n=0$ modes, which give infinite
light-cone energy, and restrict to positive integer $n$. This
regularization was also shown to preserve supersymmetry \cite{MSS95}.
We introduce discretized oscillators $$ A_{ij} (n)={1\over \sqrt{L}}
a_{ij} (k=n/L),\qquad B_{ij} (n)={1\over \sqrt{L}} b_{ij} (k=n/L), $$
which satisfy commutation relations
\mathld{
[A_{ij}(n),A_{kl}^{\dagger}(n')] & = &
\delta_{nn'}\delta_{ik}\delta_{jl} \nonumber \\
\{ B_{ij}(n),B_{kl}^{\dagger}(n') \} & = &
\delta_{nn'}\delta_{ik}\delta_{jl}.
\label{eq.commutation}}
The discretized expressions for the supercharges are
\mathld{
i Q_- & = & \sum_{n=1} \sqrt{n} \left(
B_{ij}^{\dagger}(n)A_{ij}(n) -
A_{ij}^{\dagger}(n)B_{ij}(n)\right)
\nonumber \\
\sqrt{2} Q_+ & = & \sum_{n=1}  \frac{1}{\sqrt{n}}\left(
B_{ij}^{\dagger}(n)A_{ij}(n)
+ A_{ij}^{\dagger}(n) B_{ij}(n)\right) \nonumber \\
& & - y \sum_{n_1=1}\sum_{n_2=1} \left\{\rule{0in}{3ex} \right.\nonumber \\
& &\ \ \ \ \ \ \ \frac{1}{\sqrt{n_1 n_2}}
\left[
A_{ij}^{\dagger}(n_1) A_{jk}^{\dagger}(n_2) B_{ik}(n_1+n_2)
+ B_{ik}^{\dagger}(n_1+n_2)A_{ij}(n_1)A_{jk}(n_2)
\right]
\nonumber \\
& &\ \ \ \ \ \ \ \frac{1}{\sqrt{n_1(n_1+n_2)}}
\left[A_{ij}^{\dagger}(n_1)B_{jk}^{\dagger}(n_2)A_{ik}(n_1+n_2)
+ A_{ik}^{\dagger}(n_1+n_2) A_{ij}(n_1) B_{jk}(n_2)\right] \nonumber \\
& &\ \ \ \ \ \ \ \frac{1}{\sqrt{n_2(n_1+n_2)}}
\left[B_{ij}^{\dagger}(n_1)A_{jk}^{\dagger}(n_2)A_{ik}(n_1+n_2)
+ A_{ik}^{\dagger}(n_1+n_2) B_{ij}(n_1) A_{jk}(n_2)\right] \left.
\rule{0in}{3ex} \right\}.  \nonumber
}

The Hilbert space of this model can be constructed by acting on the
Fock vacuum with mode creation operators $A_{ij}^{\dagger}$ and
$B_{ij}^{\dagger}$. As in ref. \cite{LightconeC=2,MoreC=2}, we
restrict our attention to singlet states,
\footnote{The non-singlet states do not have an obvious
stringy interpretation. As discussed in ref. \cite{LightConeGauged}
they may be eliminated by adding some amount of the gauge interaction.}
i.e. states of the form
$$ | \Phi \rangle \sim \tr \left[ A^{\dagger}(n_1) A^{\dagger}(n_2)
B^{\dagger}(n_3) \cdots A^{\dagger}(n_i) \right] |0\rangle. $$
The light-cone momentum of such a state is given by $P_-=K/L$, where
$K=\sum_i n_i$, and we restrict ourselves to states of fixed $P_-$.
The integer $K$ plays the role of the cut-off and, remarkably,
for a finite $K$ the total number of states is finite  \cite{PauliBrodsky}.
The continuum limit is recovered as $L$ and $K$ are sent to infinity.
For the sake of illustration, we display the set of
states with $K = 3$. There are five bosonic states
\mathld{
|1\rangle_b & = & \frac{1}{N^{3/2}\sqrt{3}} \tr\left[A^\dagger(1)
A^{\dagger}(1) A^{\dagger}(1) \right] | 0 \rangle \nonumber \\
|2\rangle_b & = & \frac{1}{N^{3/2}}\tr\left[A^\dagger(1) B^{\dagger}(1)
B^{\dagger}(1) \right] | 0 \rangle \nonumber \\
|3\rangle_b & = & \frac{1}{N}\tr\left[A^\dagger(2) A^{\dagger}(1) \right] | 0
\rangle \nonumber \\
|4\rangle_b & = & \frac{1}{N} \tr\left[B^\dagger(2) B^{\dagger}(1) \right] | 0
\rangle \nonumber \\
|5\rangle_b & = & \frac{1}{N^{1/2}}\tr\left[ A^{\dagger}(3) \right] | 0 \rangle
\nonumber
}
and five fermionic states
\mathld{
|1\rangle_f & = & \frac{1}{N^{3/2}} \tr\left[A^\dagger(1) A^{\dagger}(1)
B^{\dagger}(1) \right] | 0 \rangle \nonumber \\
|2\rangle_f & = & \frac{1}{N^{3/2}\sqrt{3}}\tr\left[B^\dagger(1) B^{\dagger}(1)
B^{\dagger}(1) \right] | 0 \rangle \nonumber \\
|3\rangle_f & = & \frac{1}{N} \tr\left[A^\dagger(2) B^{\dagger}(1) \right] | 0
\rangle \nonumber \\
|4\rangle_f & = & \frac{1}{N} \tr\left[B^\dagger(2) A^{\dagger}(1) \right] | 0
\rangle \nonumber \\
|5\rangle_f & = & \frac{1}{N^{1/2}}\tr\left[ B^{\dagger}(3) \right] | 0
\rangle. \nonumber \\
\label{eq.supercharge.modes}
}
For higher values of $K$ all states may be generated systematically by
considering all partitions of $K$ into positive integers, assigning
one of the two oscillators, $A^\dagger$ or $B^\dagger$, to each
integer, and eliminating null and redundant states. Once the basis of
states for a given $K$ has been found, it is straightforward to
determine the matrix elements of $I$.

As in the bosonic case, the mode operators create or annihilate
``string bits,'' except now there are two species of such bits. The
stringy states we are interested in may be thought of as bound states
of these string bits. Various terms in the supercharges implement
joining and splitting of neighboring bits or an interchange of a
bosonic and a fermionic bits.

In what follows, we describe the result of our analysis for the
spectrum of $I = \sqrt{2} i Q_+Q_-$. We constructed the exact matrix
representation of $I$ symbolically for finite $K$ and then evaluated
its spectrum numerically.  The goal is to examine the convergence of
the spectrum in the large $K$ limit in order to recover the continuum
theory.  Unfortunately, the dimension of matrix $I$ grows rapidly with
$K$.  Our computer resources have allowed us to perform
diagonalizations for $K = $ 3, 4, 5, 6, 7, 8, 9, and 10, where the
dimensions of the matrix $I$ are 5, 10, 25, 62, 157, 410, 1097, and
2954, respectively.  The number of states for each $K$ is greater than
that found in \cite{MSS95} by one, since our model is not gauged and
we include the single string bit state.

\begin{figure}[htb]
\centerline{\psfig{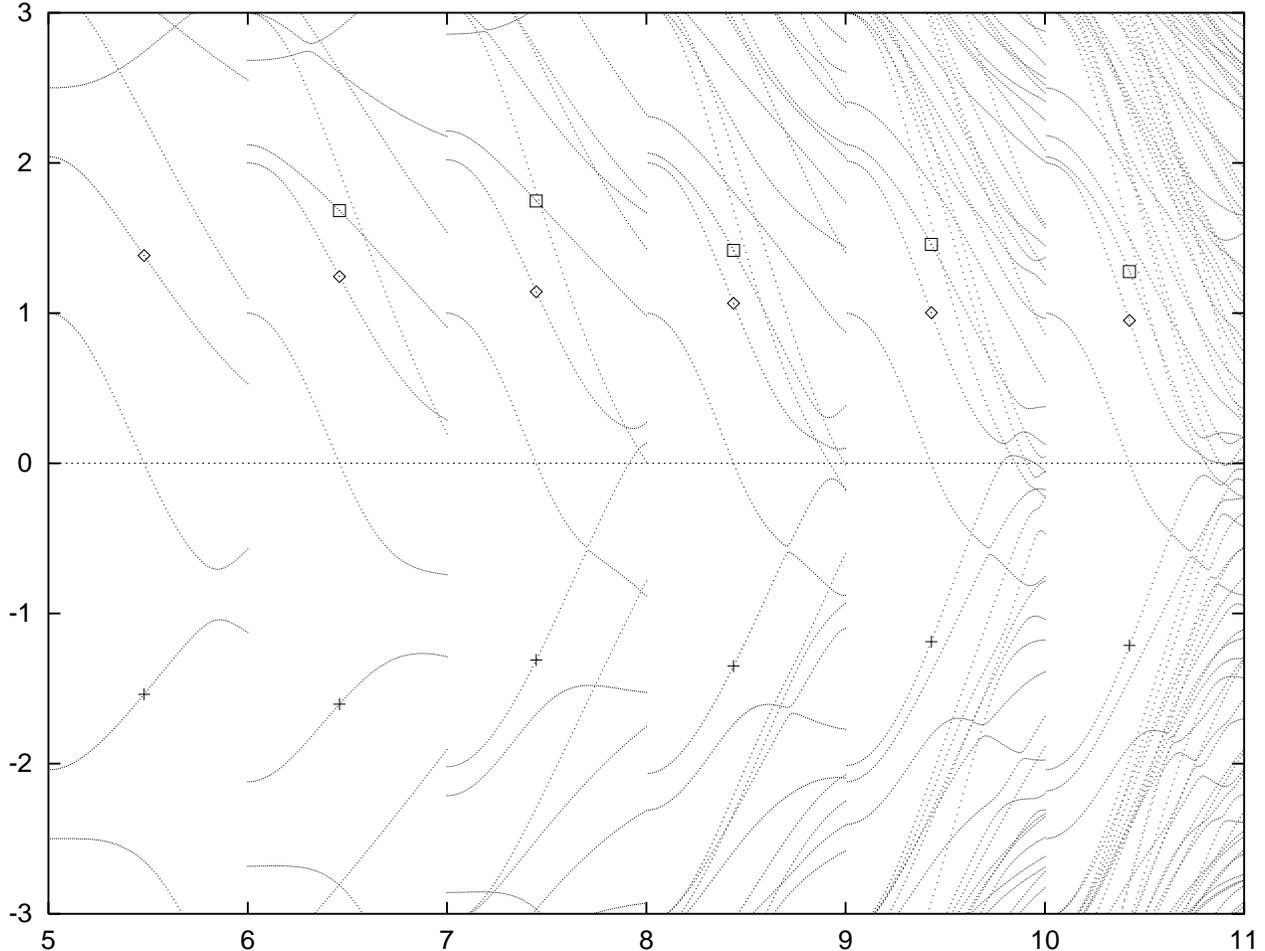}}
\caption{Low lying spectrum of $I$  for $0<y<1$ for $K$ from{} 5 to 10}
\label{fig.fullspectrum}
\end{figure}

In figure
\ref{fig.fullspectrum}
we plot the spectrum of $I$ as a function of $y$ for $K$ ranging form
5 to 10 (we show only the eigenvalues whose magnitude does not exceed
3).
Note that, for each $K$, there is a special value of $y$ where the
first zero eigenvalue appears. It is plausible that this phenomenon
persists to arbitrarily large values of $K$, and that this zero
crossing signifies a phase transition.  Indeed, we found the locations
of these zero crossings apparently converging like a power with
increasing $K$. We have calculated these locations and tried a least
squares fit to the functional form $a + b/K^c$. This gives an estimate
of $y_c=0.37$ for the zero crossing of the first level in the large
$K$ limit.

\begin{figure}[htb]
\centerline{\psfig{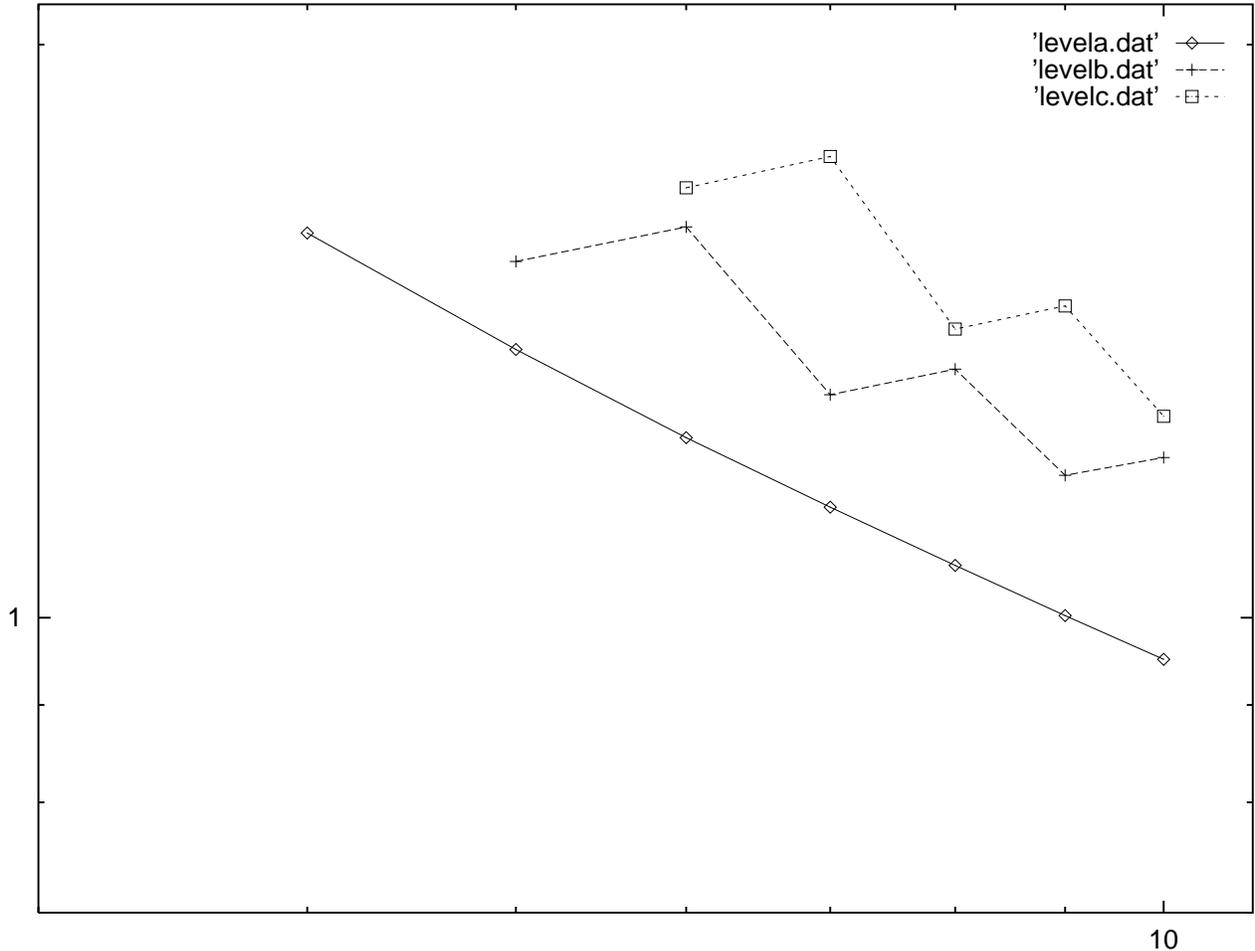}}
\caption{Spectrum 2nd, 3rd, and 4th levels at the zero of 1st level v.s. $K$}
\label{fig.lowlevels}
\end{figure}

We are also interested in the behavior of other states with low
mass-squared.  If a continuum of states is forming as we conjectured,
the $n$-th level for any finite $n$ should develop a zero eigenvalue
at $y=y_c$ in the limit $K \rightarrow \infty$.  This behavior is not
at all obvious from{} figure \ref{fig.fullspectrum}.  To examine the
situation more closely, we found the eigenvalues of other low lying
states for the value of $y$ which corresponds to the zero crossing of
lowest state and studied their dependence on $K$. These points are
marked on figure \ref{fig.fullspectrum} by tick marks on the spectrum.
We demonstrate the $K$ dependence of these quantities in figure
\ref{fig.lowlevels}.
Figure \ref{fig.lowlevels} is plotted on a log-log scale in
anticipation of the power law dependence.  The linearity of the graph
for the second level appears to confirm this.  The third and fourth
level also appear to follow this trend superposed on some
oscillations. The oscillation is likely due to the difference between
the regularizations with odd and even $K$.  For high enough $K$ this
oscillation should be suppressed and these low lying states will
hopefully exhibit a power law decay similar to the second level.
Alternatively we could extrapolate for even and odd $K$ separately,
but we do not have enough data for that.

\begin{figure}[htb]
\centerline{\psfig{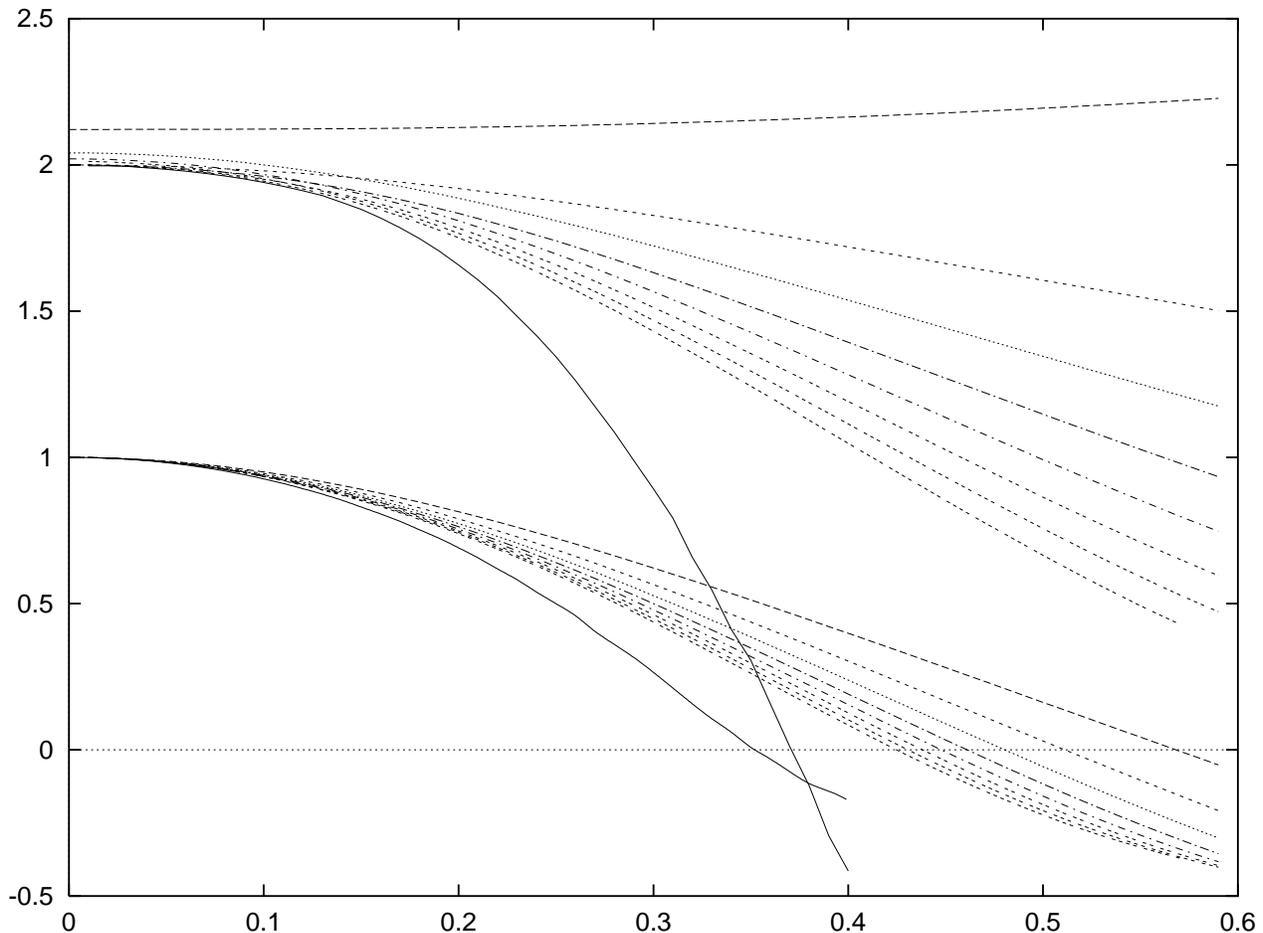}}
\caption{Spectrum of two lightest states for various $K$'s and their
extrapolation.  Dotted lines are for $K$ from{} 3 to 10.  The solid line
is the extrapolation.}
\label{fig.extrapolate}
\end{figure}

As a final check, we concentrated on the two lowest states and
attempted to extrapolate their masses to infinite $K$ for each $y$.
Again, the spectrum appeared to be converging like a power.  Using
best fits to the form $a+b/K^c$, we determined our best guesses for
the large $K$ limit of the two lowest masses.  The result of this
calculation is illustrated in figure \ref{fig.extrapolate} where we
plot the dependence on $y$ of the two lowest masses (we show the
results for fixed $K$ as well as the extrapolations to infinite $K$).
We are encouraged to find that the extrapolated masses for the two
lowest states come very close to vanishing simultaneously at $y$ near
0.37, which is also the value of $y_c$ that we found by extrapolating
the locations of the zero crossings.

At the moment our results are far from{} being thoroughly convincing,
since our extrapolations are no more than the best guesses we can
offer based on the available computational data.  With the computer
resources available to us we were unable to extend the calculation to
$K>10$. Bigger computers, or improved algorithms, should certainly
permit the extension of our calculations to higher $K$. It would
certainly be interesting to check the continuation of the trends we
observed and perform more reliable extrapolations.

In conclusion, we have presented some numerical evidence which
suggests that the spectrum of a supersymmetric large-$N$ matrix field
theory becomes continuous and massless at a critical value of the
coupling constant, $\lambda_c$.  A similar phenomenon occurs in the
$c=1$ matrix model where $\lambda_c$ corresponds to the onset of the
world sheet continuum limit, and the Liouville mode emerges as an
additional target space coordinate.  By analogy, we speculate that our
1+1 dimensional matrix model describes a 2+1 dimensional non-critical
superstring in some background.  If true, one might further consider
an exciting possibility of finding a non-perturbative matrix model
formulation of higher dimensional superstring theories.

By construction, the conjectured 2+1 theory will possess (1,1)
supersymmetry in two of its dimensions.  From{} the special nature of
the Liouville mode, however, we find it extremely unlikely that this
symmetry is extended to 2+1 dimensions.  As such, it is similar in
spirit to the model considered in \cite{SK90}.  It would be
interesting to further explore the relationship between these two
models. Perhaps more can be learned about the relation between our
approach and its world sheet formulation by studying the
renormalization group flow and the operator algebra of the world sheet
action (\ref{eq.worldsheet}) along the lines of \cite{CP94}.  It would
also be interesting to explore the relation to the Green-Schwartz
approach taken in \cite{SiegelGS,SiegelLiouville}.  The NSR approach,
the Green-Schwartz approach, and the matrix models offer complementary
methods for the definition of supersymmetric non-critical strings.  It
would be very interesting to find a model which we understand in all
three formulations.

\vspace{2ex}

\noindent {\Large{\bf Acknowledgement}}\\

\noindent We would like to thank Shyamoli Chaudhuri and Kresimir
Demeterfi for discussions.  This work was supported in part by DOE
grant DE-FG02-91ER40671, the NSF Presidential Young Investigator Award
PHY-9157482, James S. McDonnell Foundation grant No. 91-48, and an
A. P. Sloan Foundation Research Fellowship.

\bibliographystyle{plain}

\end{document}